# Concentrating Electric and Thermal Fields Simultaneously Using Fan-shaped Structure

Chuwen Lan[1,2], Bo Li[2]*, Ji Zhou[1]*

[1]State Key Laboratory of New Ceramics and Fine Processing, School of Materials Science and Engineering, Tsinghua University, Beijing 100084, China

[2]Advanced Materials Institute, Shenzhen Graduate School, Tsinghua University, Shenzhen, China

*Corresponding author: zhouji@mail.tsinghua.edu.cn, libo@mail.tsinghua.edu.cn

**Abstract:**

Recently, considerable attention has been focused on the transformation optics and metamaterial due to their fascinating phenomena and potential applications. Concentrator is one of the most representative ones, which however is limited in single physical domain. Here we propose and give the experimental demonstration of bifunctional concentrator that can concentrate electric and thermal fields into a given region simultaneously while keeping the external fields undistorted. Fan-shaped structure composed of alternating wedges made of two kinds of natural materials is proposed to achieve this goal. The simulation and experimental results show good agreement, thereby confirming the feasibility of our scheme.

**Main Text**

With its fascinating phenomena and potential applications like negative index refraction[1], perfect absorber[2] and invisibility cloak[3-5], metamaterial (MM) is currently drawing considerable attention. Another key motivation is the well-developed transformation optics (TO), which is based on the fact that physical equation form is invariant under the coordinate transformation. MMs, with combination of TO, have paved a novel way towards manipulation of physical field in a desired way. Motived by their great progress made in electromagnetic wave[3-9], MM and TO are rapidly extended to other waves like matter wave[10], acoustic wave[11] and elastic wave[12]. Most recently, MM and TO are successfully employed to manipulate scalar fields like static magnetic field [13-14], current field[15-16], thermal field[17-22], diffusive mass[23-25] and electrostatic field[26].

Although great achievement has been made in manipulation of various fields, they are usually limited in single physical domain. Recently, bifunctional metamaterials and transformation multiphysics were proposed to achieve independent and simultaneous manipulation of multi-physics field[27]. The first experimental realization of manipulation of multi-physics field was then reported to cloak the thermal and electric fields simultaneously using artificial composite[28]. Our group has proposed and given the experimental demonstration of simultaneous and independent manipulation of electric and thermal fields using a bilayer structure[29]. Just recently, we have proposed a general method to obtain simultaneous manipulation of multi-physics field with combination of passive and active schemes[30]. Up to now, however, no work has been reported on the bifunctional concentrator that can concentrate multi-physics field into a given region simultaneously.

Here, we propose a bifunctional concentrator that can concentrate electric and thermal fields into a given region simultaneously while keeping the external fields undistorted. We further fabricated and tested it. The simulation and experimental results show good agreement, directly confirming the feasibility of our scheme.

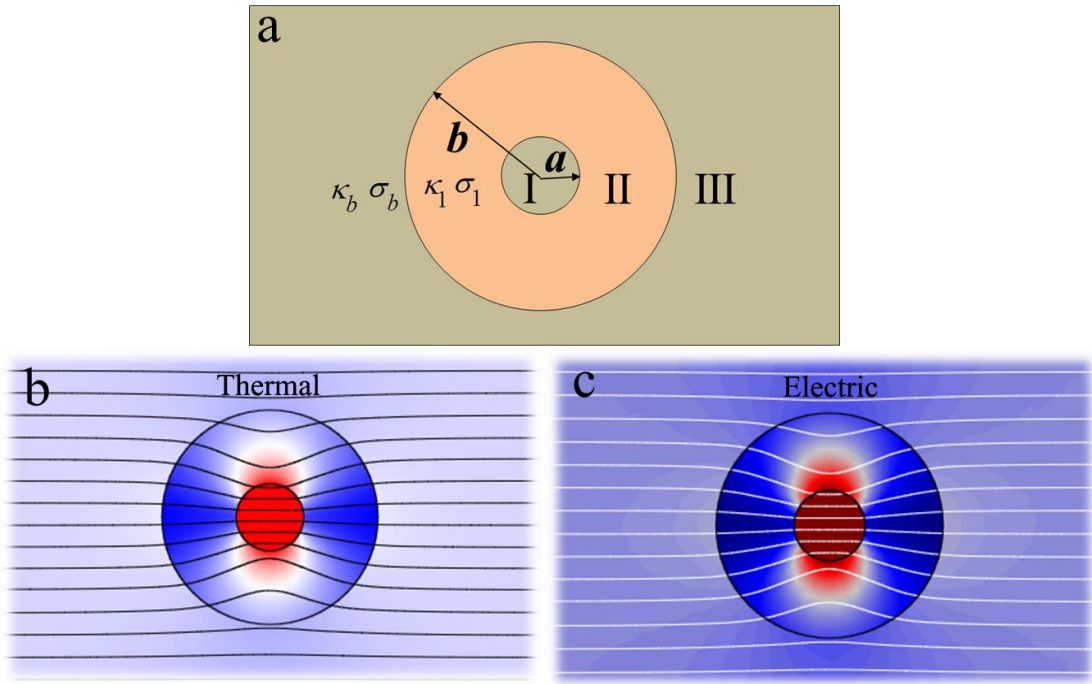

Figure 1. The principle for bifunctional concentrator: a) The corresponding physical model. b) The thermal flux distribution. c) The electric current distribution.

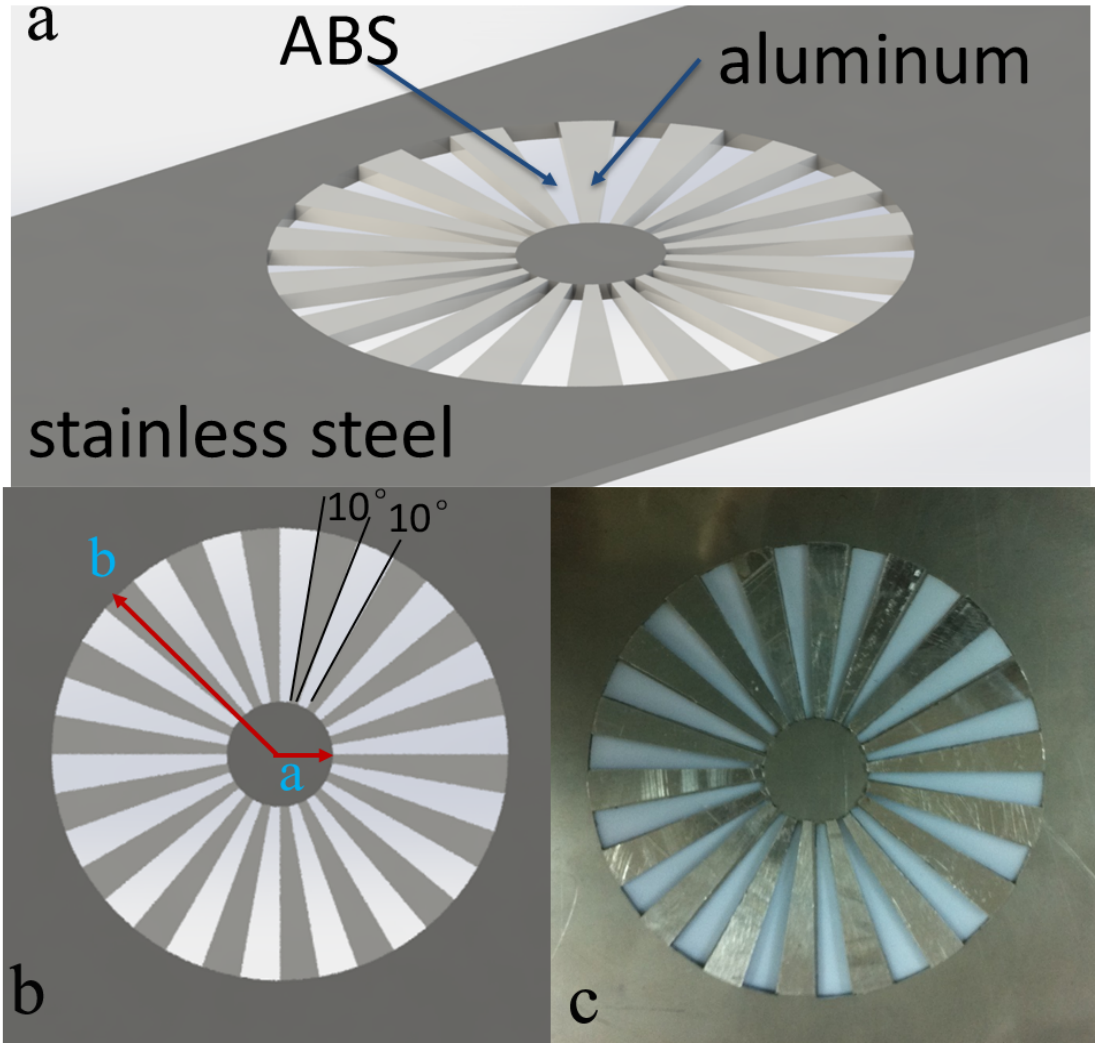

**Figure 2.** (a) The schematic illustration for practical realization of bifunctional concentrator. (b) The geometrical parameters: a=6mm, b=30mm. (c)The photograph of fabricated sample.

Suppose that the space is divided into three parts: core region ($r<a$), shell region ($a<r<b$) and exterior region ($r>b$). The core region and exterior region are made of background material with thermal conductivity $\kappa_0$ and electric conductivity $\sigma_0$. According to the previous work[17], to make a thermal concentrator, the thermal conductivity of shell should be anisotropic and satisfy this relationship: $\kappa_\theta \kappa_r = \kappa_0^2$, where $\kappa_r > \kappa_\theta$ ($\kappa_r$ and $\kappa_\theta$ are the thermal conductivity components of shell in radial and circular directions, respectively). Similarly, as for electric concentrator, the electric conductivity of shell should satisfy this relationship: $\sigma_\theta \sigma_r = \sigma_0^2$, where $\sigma_r > \sigma_\theta$ ($\sigma_r$ and $\sigma_\theta$ are the electric conductivity components of shell in radial and circular directions, respectively). Clearly, it is challenging to implement a shell satisfying these two conditions simultaneously. To accomplish this goal, one may employ bifunctional metamaterial[27,28], which however suffers from complex fabrication. To overcome this problem, we utilize so-called fan-shaped structure as shown in Figure 2, which is composed of alternating 18 wedges made of material A ( ABS: $\kappa_A$=0.15 W/mK, $\sigma_A$ =0 S/m) and 18 wedges made of material B ($\kappa_B$ and $\sigma_B$). According to effective medium theory[31], the thermal and electric conductivity components can be expressed as:

$$\kappa_\theta = \frac{1}{\frac{1}{\kappa_A} \times f_1 + \frac{1}{\kappa_B} \times f_2} = \frac{1}{\frac{1}{0.3} + \frac{1}{2\kappa_B}} \tag{1}$$

$$\kappa_r = \kappa_A \times f_1 + \kappa_B \times f_2 = 0.075 + 0.5 \times \kappa_B \tag{2}$$

$$\sigma_\theta = \frac{1}{\frac{1}{\sigma_A} \times f_1 + \frac{1}{\sigma_B} \times f_2} = 0 \tag{3}$$

$$\sigma_r = \sigma_A \times f_1 + \sigma_B \times f_2 = 0.075 + 0.5 \times \sigma_B \tag{4}$$

Here, assume that the background material is made of stainless steel with $\kappa_0$=15 W/mK and $\sigma_0$=1.4e6 S/M. Clearly, to obtain a bifunctional concentrator, one should make $\kappa_B > \kappa_0$ and $\sigma_B > \sigma_0$. Then, one can obtain: $\kappa_\theta = 0.3$, $\kappa_r \approx 0.5\kappa_B$, $\sigma_\theta = 0$, $\sigma_r \approx 0.5\sigma_B$. Apparently, one can see that it is hard to find material B to satisfy the aforementioned conditions. Fortunately, these two conditions can be simplified. Simulations were carried out to investigate the concentrator performance with various $\kappa_r$ and $\kappa_\theta$ by making $\kappa_\theta / \kappa_0 = 0.3 / \kappa_0 = 0.02$, $\sigma_\theta / \sigma_0 = 0$. The thermal concentrator performance is shown in Figure 3a-i, where one can find when $\kappa_r / \kappa_0 > 6$, a nearly perfect thermal concentrator can be obtained. Similarly, the case for electric field is shown in Figure 4a-i, where one can find that when $\sigma_r / \sigma_0 > 6$, a nearly perfect electric concentrator can be obtained. Thereby, to obtain a bifunctional concentrator, one should make $\kappa_B > 12\kappa_0 = 180$ W/mK, $\sigma_B > 12\sigma_0 = 2.4e7$ S/M. This is an important property, which can greatly simplify the fabrication and many natural materials are available. As for our study, aluminium, silver, copper can be used. Here, we chose aluminium with thermal conductivity $\kappa_B$ =237 W/mK and electric conductivity $\sigma_B$=3.57e7 S/M.

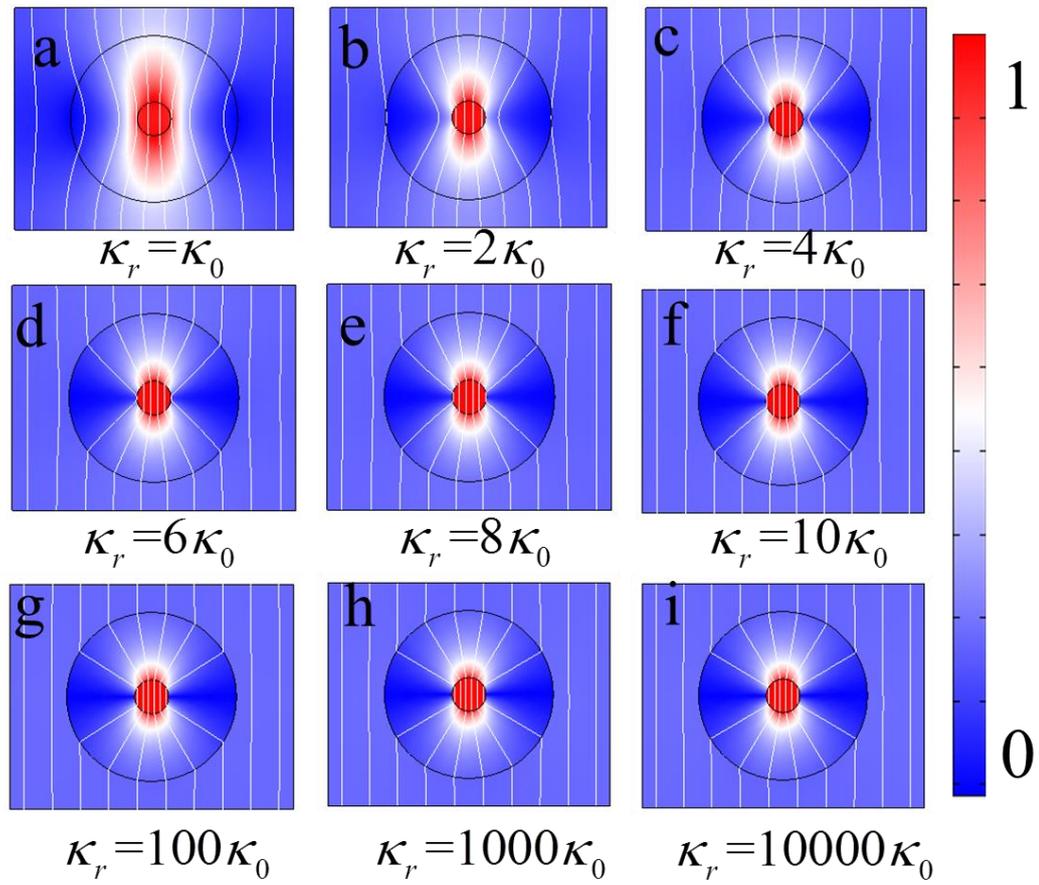

Figure 3. Thermal flux distribution for different $\kappa_r$ values: (a) $\kappa_r=\kappa_0$, (b) $\kappa_r=2\kappa_0$, (c) $\kappa_r=4\kappa_0$,(d) $\kappa_r=6\kappa_0$, (e) $\kappa_r=8\kappa_0$ (f) $\kappa_r=10\kappa_0$ (g) $\kappa_r=100\kappa_0$ (h) $\kappa_r=1000\kappa_0$. The white lines represents isothermal lines.

To confirm our prediction, simulations based on Multiphysics finite element soft, Comosol, were carried out to obtain its thermal and electric conductivity properties. Firstly, we provide the simulated results for the homogeneous background material (see Figure 5a-b). As shown in these pictures, a uniform temperature gradient is generated from high temperature (80 ℃) to low temperature (0 ℃), meanwhile, uniform electric potential gradient is produced from high potential (1V) to low potential (0V). The results for our bifunctional device are shown in Figure 5c-d, where both thermal and electric fields are concentrated into the core region resulting enhancement for temperature and electric potential gradient, meanwhile, the external fields are nearly undistorted, thereby indicating good thermal and electric concentrator performance. Note that the slight distortion for thermal field and electric field can be attributed to the discrete of fan-shaped structure.

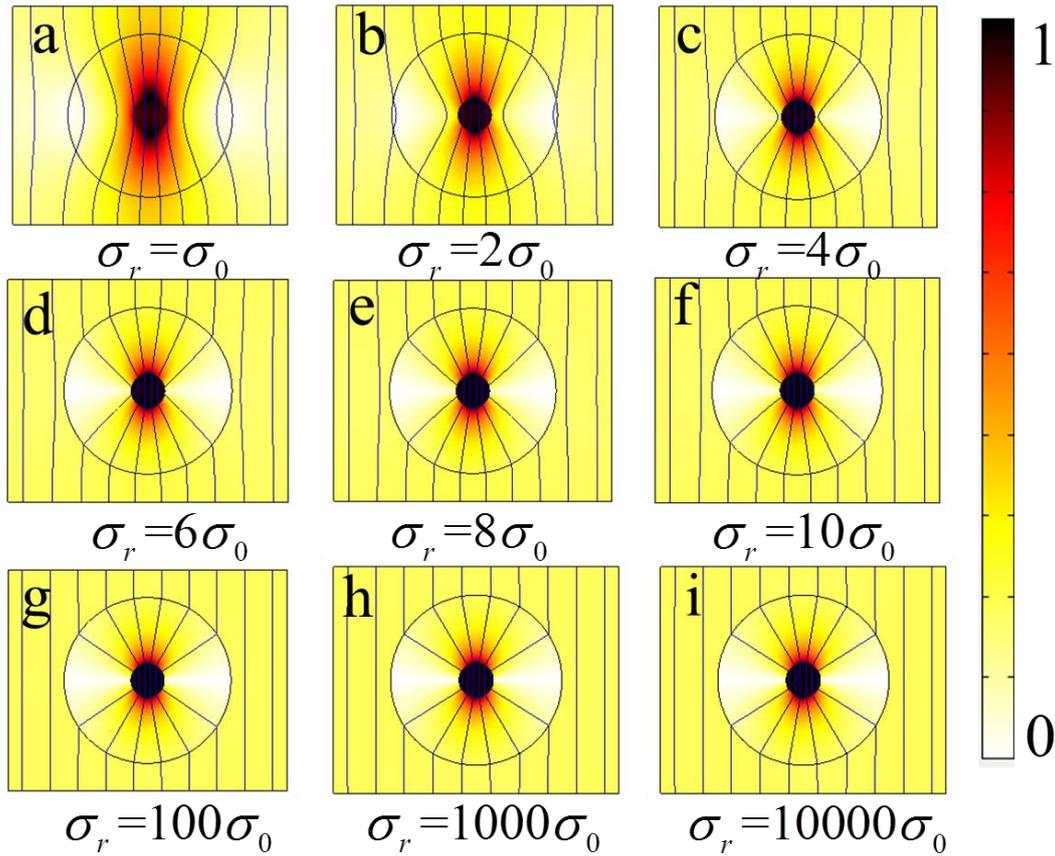

Figure 4. Current density distribution for different $\sigma_r$ value: (a) $\sigma_r=\sigma_0$, (b) $\sigma_r=2\sigma_0$, (c) $\sigma_r=4\sigma_0$, (d) $\sigma_r=6\sigma_0$, (e) $\sigma_r=8\sigma_0$ (f) $\sigma_r=10\sigma_0$ (g) $\sigma_r=100\sigma_0$ (h) $\sigma_r=1000\sigma_0$. The white lines represent isopotential lines.

Experimentally, such device was fabricated and shown in Figure 3c. In the measurement, the two sides of sample touched hot water (80 ℃) and ice water (0 ℃), respectively. An infrared heat camera (Fluke Ti300) was used to obtain the temperature profile distribution. After several minutes, the steady temperature profile is generated and shown in Figure 6. Clearly, the temperature gradient in the core region is enhanced and the external field is nearly undistorted, thus suggesting good thermal concentrator performance. To evaluate the electric concentrator performance, according to the previous work[16], one can employ gradual structure and measure the potential distribution along lines x=31mm, x=-31mm and y=0mm (see inserts in Figure 7a-c). As respected, the potential distribution along the lines x=31mm and x=-31mm are straight, which means no distortion. Meanwhile, the potential gradient in the core region is apparently enhanced. The measurement results are shown in Figure 7, indicating good agreement with the simulated ones, thus suggesting a good electric concentrator.

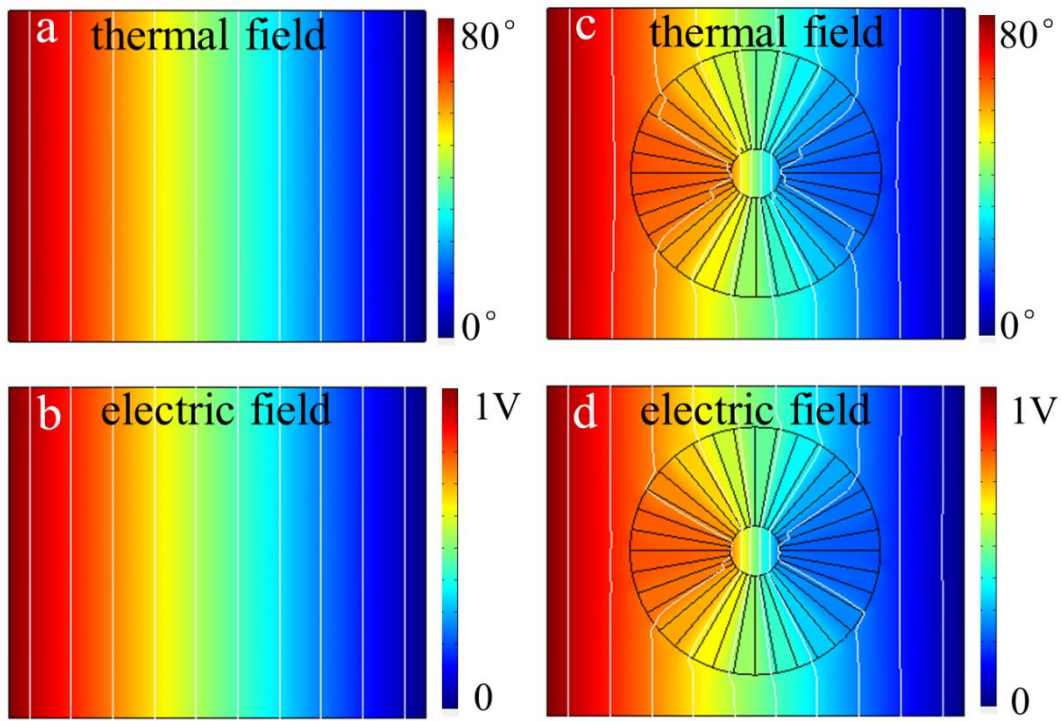

**Figure 5.** Simulation results. (a) Temperature profile for homogeneous background material. (b) Electric potential distribution for homogeneous background material. (c) Temperature profile for bifunctional concentrator. (d) Electric potential distribution for bifunctional concentrator.

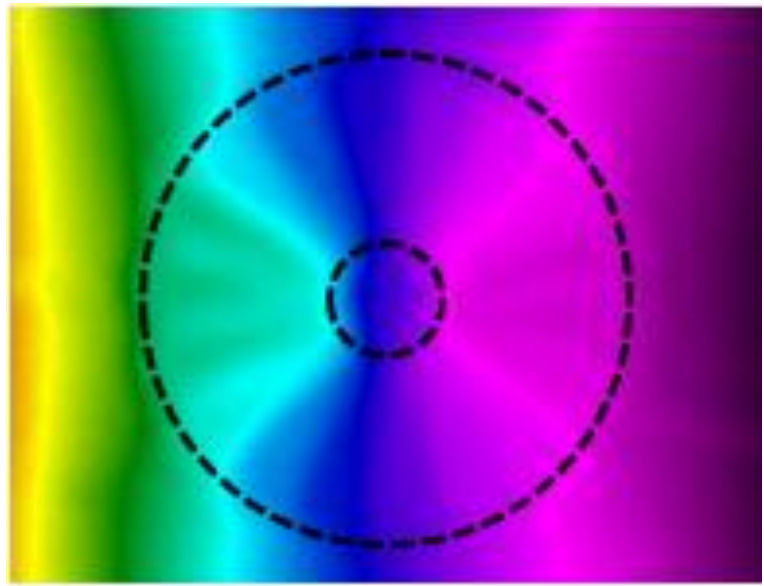

Figure 6. Experimental measured temperature profile for the bifucntional device. The black circles represent the sample.

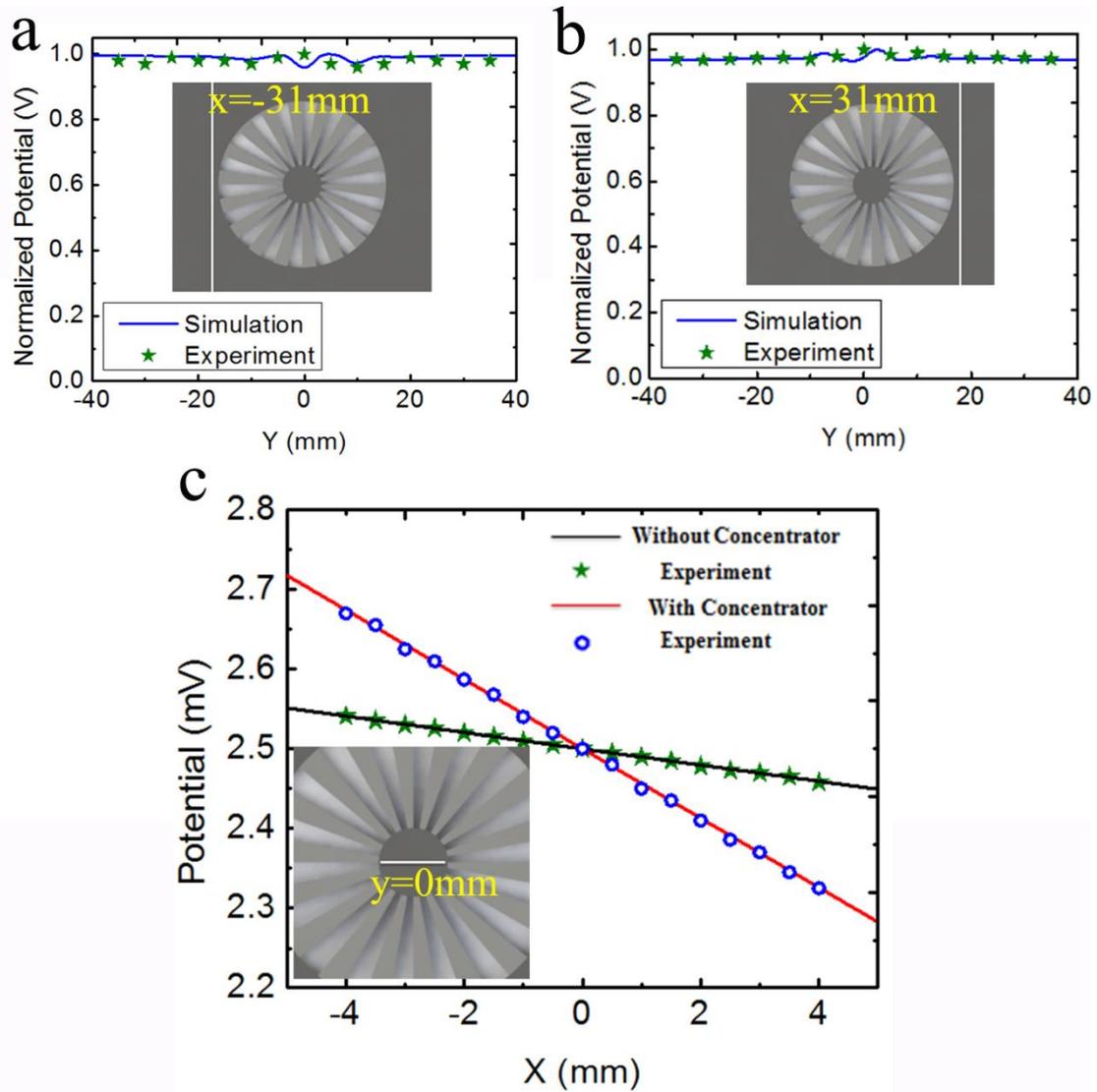

**Figure 7.** The simulation and experiment results for the different cases at corresponding positions: (a) x=-31mm, (b) x=31mm and (c) y=0mm. The white lines in inserts represent observed lines.

In conclusion, we have proposed and given the experimental demonstration of bifunctional concentrator that can concentrate electric and thermal fields into a given region simultaneously while keeping the external fields undistorted. Here, only natural materials are used, thus greatly simplifying the fabrication. The simulation and experimental results show good agreement, thereby confirming the feasibility of our scheme. This is a general method for achieving bifunctional concentrator, which can also be extended to other multi-physics systems.

This work was supported by the National Natural Science Foundation of China under Grant Nos. 51032003, 11274198, 51221291 and 61275176, National High Technology Research and

Development Program of China under Grant No. 2012AA030403, Beijing Municipal Natural Science Program under Grant No. Z141100004214001, and the Science and technology plan of Shenzhen city under grant Nos.JCYJ20120619152711509, JC201105180802A and CXZZ20130322164541915.